\documentclass[aip,preprint,showpacs,showkeys,superscriptaddress]{revtex4-1}

\usepackage{amssymb,amsmath,graphicx}
\begin{document}

\title{Surface acoustic wave controlled carrier injection into self-assembled quantum dots and quantum posts}
\author{Hubert J. Krenner}
\email{hubert.krenner@physik.uni-augsburg.de}
\affiliation{Lehrstuhl f\"ur Experimentalphysik 1 and ACIT, Universit\"at Augsburg, Universit\"atsstr. 1, 86159 Augsburg, Germany}

\author{Stefan V\"olk}
\affiliation{Lehrstuhl f\"ur Experimentalphysik 1 and ACIT, Universit\"at Augsburg, Universit\"atsstr. 1, 86159 Augsburg, Germany}
\author{Florian J. R. Sch\"ulein}
\affiliation{Lehrstuhl f\"ur Experimentalphysik 1 and ACIT, Universit\"at Augsburg, Universit\"atsstr. 1, 86159 Augsburg, Germany}
\author{Florian Knall}
\affiliation{Lehrstuhl f\"ur Experimentalphysik 1 and ACIT, Universit\"at Augsburg, Universit\"atsstr. 1, 86159 Augsburg, Germany}
\author{Achim Wixforth}
\affiliation{Lehrstuhl f\"ur Experimentalphysik 1 and ACIT, Universit\"at Augsburg, Universit\"atsstr. 1, 86159 Augsburg, Germany}
\author{Dirk Reuter}
\affiliation{Lehrstuhl f\"ur Angewandte Festk\"orperphysik, Ruhr-Universit\"at Bochum, Universit\"atsstr. 150, 44780 Bochum, Germany}
\author{Andreas D. Wieck}
\affiliation{Lehrstuhl f\"ur Angewandte Festk\"orperphysik, Ruhr-Universit\"at Bochum, Universit\"atsstr. 150, 44780 Bochum, Germany}
\author{Tuan A. Truong}
\affiliation{Materials Department, University of California, Santa Barbara CA 93106, United States}
\author{Hyochul Kim}
\affiliation{Materials Department, University of California, Santa Barbara CA 93106, United States}
\affiliation{Department of Electrical and Computer Engineering, IREAP, University of Maryland, College Park, MD 20742, United States}
\author{Pierre M. Petroff}
\affiliation{Materials Department, University of California, Santa Barbara CA 93106, United States}

\begin{abstract}%

We report on recent progress in the acousto-electrical control of self-assembled quantum dot and quantum post using radio frequency surface acoustic waves (SAWs). We show that the occupancy state of these optically active nanostructures can be controlled via the SAW-induced dissociation of photogenerated excitons and the resulting \emph{sequential} bipolar carrier injection which strongly favors the formation of neutral excitons for quantum posts in contrast to conventional quantum dots.
We demonstrate high fidelity preparation of the neutral biexciton which makes this approach suitable for deterministic entangled photon pair generation. The SAW driven acoustic charge conveyance is found to be highly efficient within the wide quantum well surrounding the quantum posts. Finally we present the direct observation of acoustically triggered carrier injection into remotely positioned, individual quantum posts which is required for a low-jitter SAW-triggered single photon source.
\end{abstract}

\maketitle  

\section{Introduction}
\begin{figure}[htb]%
%  \sidecaption
  \includegraphics*[width=.9\columnwidth]{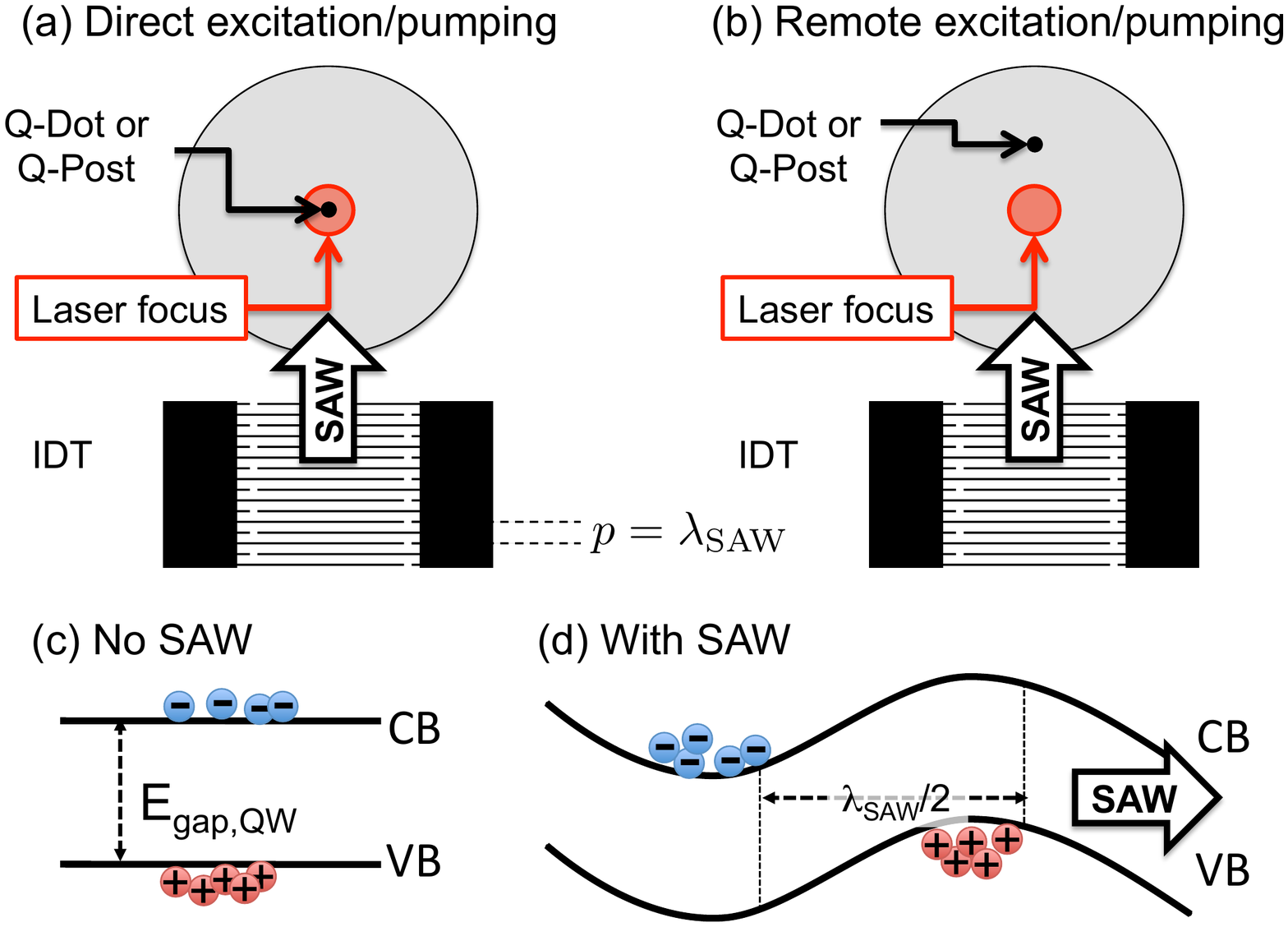}
  \caption[]{%
   (color online) (a,b) Direct and remote excitation of a single Q-Dot/Q-Post nanoemitter. Band edge energies without (c) and with a SAW applied (d). The Type-II bandedge modulation by the SAW gives rise to exciton dissociation into individual electrons and holes, suppressing radiative recombination.}
    \label{directandremote}
\end{figure}

Surface acoustic waves (SAW) have proven to be a powerful tool for the control of optically active nanostructures at radio frequencies. On compound III-V semiconductors such as GaAs or InP these SAWs can be generated all-electrically by employing lithographically defined interdigital transducers (IDTs). In this scheme, which is depicted in Fig.~\ref{directandremote} (a), a RF voltage is applied to the IDT and if the periodicity $p= \lambda_{\rm SAW}$ matches the dispersion relation $f_{\rm RF}=c_{\rm s} /\lambda_{\rm SAW}$, with $c_{\rm s}$ being the sound velocity, a SAW is resonantly generated. 
In an optical experiment individual nanostructures can be studied under the influence of a SAW. The strain fields of the SAW induce large piezoelectric fields in III-V semiconductors which have a strong impact on the optical properties. As shown in Fig.~\ref{directandremote} (c) and (d), the band edges are flat for no SAW and Type-II modulated with a SAW applied. The resulting spatial separation of photogenerated electron-hole pairs gives rise to a pronounced and dynamic suppression of the optical emission of quantum wells (QWs)~\cite{Rocke,Alsina}. Moreover, a SAW propagates with $c_{\rm s}$ on the crystal surface and, therefore, these carriers are acoustically transported in a conveyor-type way in the plane of the QW~\cite{Rocke}. This property sparked the idea to use this mechanism to realize an precisely triggered single photon source~\cite{Wiele}. This remote acoustic pumping of a single quantum dot (Q-Dot) [c.f. Fig.~\ref{directandremote} (d)] was first demonstrated for an ensemble~\cite{Boedefeld} and very recently also for single interface fluctuation Q-Dots~\cite{Couto}. Here we present first experiments investigating the potential of the most advanced semiconductor system, self-assembled Q-Dots and a novel type of nanostructure, self-assembled quantum posts (Q-Posts)~\cite{He_NanoLett,Krenner_PhysicaE,Krenner-review,Ridah_Review} which are both found to be high quality single photon emitters~\cite{Michler_Science,Krenner_NanoLett}. We show that the SAW-induced dissociation of excitons and the resulting sequential injection of electrons and holes can be used to (i) enhance the carrier injection efficiency and (ii) control the charge state of the nanostructure. Due to their unique morphology, Q-Posts are a particularly suitable system to implement an efficient acoustically triggered single photon source and first results towards this goal are presented.

\section{Experimental results}

\subsection{Sample design and experimental setup}
We studied our samples by micro-photoluminescence (${\mu}$-PL) at low temperatures ($T=4\mathrm{K}$) using a pulsed and externally triggered diode laser ($\tau\mathrm{_{laser}<100~ps}$, $\lambda\mathrm{_{laser}=661~nm}$) focused to a $\sim2~\mu\mathrm{m}$ diameter spot to photogenerate charge carriers. The emission is spectrally analyzed with a resolution $<150~\mu\mathrm{eV}$ using a 0.5m grating monochromator and a Si-CCD array which can be used for multi-channel spectral analysis or for direct imaging of the emission over the entire field of view of the microscope objective [grey shaded in Fig.~\ref{directandremote} (a) and (b)].\\
The two heterostructures have been grown by molecular beam epitaxy (MBE) and contain a nominally undoped single layer of either self-assembled Q-Dots or height-controlled, self-assembled Q-Posts (for details see Refs. \cite{He_NanoLett,Krenner_PhysicaE,Krenner-review,Voelk_NanoLett}). These two types of nanostructures studied differ strongly in their morphology and the adjacent two-dimensional QW system they are coupled to and which can be used for acoustic charge conveyance. Conventional Q-Dots are flat, $\sim 3-5$~nm high, Indium-rich islands with a diameter in the range of 20~nm which nucleate on a thin $<1{~\rm nm}$ InGaAs wetting layer (WL) resembling a narrow and disordered QW. In contrast, self-assembled Q-Posts are columnar structures with similar diameters as the Q-Dots. In particular, their height can be adjusted with nanometer precision from $\sim3$ to $>60$~nm while preserving the high optical quality of the established Q-Dots. The Q-Posts themselves are embedded in an $\mathrm{In_{0.1}Ga_{0.9}As}$ Matrix-QW of the same thickness. The enhanced In-content within the Q-Post of $\sim40$\% ensures full, quasi-zero-dimensional confinement for electrons and holes in the Q-Post. The Q-Posts studied in these experiments are fabricated by eight repetitions of the employed deposition sequence and have a nominal height of 23~nm. For both samples, we realized a surface density gradient of the Q-Dots or Q-Posts by intentionally stopping the substrate rotation during the deposition of these layers. The position at which the transition to low surface density occurs was determined by conventional scanning ${\mu}$-PL at low temperatures for each wafer and we aligned IDTs such that the region with less than one nanostructure per $\mu\mathrm{m}^2$ was in the propagation direction of the SAW. For our experiments, we used IDTs with wavelengths of $\lambda_1 = 11.6~\mu\mathrm{m}$ for the Q-Dot sample and  $\lambda_2 = 15~\mu\mathrm{m}$ for the Q-Post sample, respectively. For the SAWs propagating along the [110] crystal directions on a (001) GaAs substrate, these wavelengths correspond to excitation frequencies of $f_1=251.5$~MHz and $f_2=193$~MHz.\\

\subsection{Direct excitation}

\begin{figure}[]%
 %\sidecaption
  \includegraphics*[width=0.9\columnwidth]{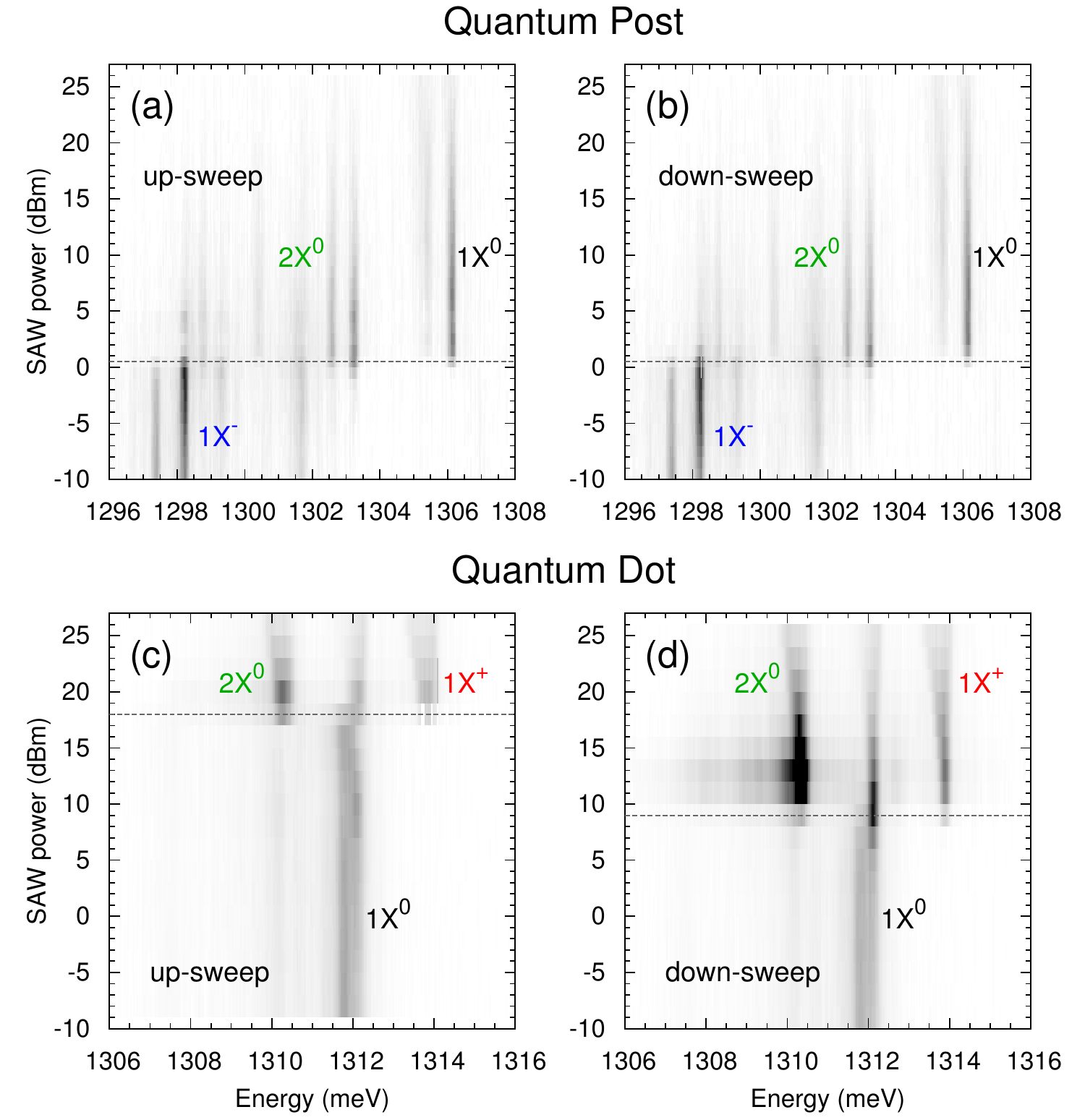}%
  \caption[]{(color online)SAW power scans of a single Q-Post (a, b) and Q-Dot (c, d) for increasing (decreasing) $P_{\rm SAW}$ on the left(right) handside panels. Both systems show a clear switching at the onset of charge conveyance which is hysteretic for the Q-Dot.}
    \label{Hysterese}
\end{figure}

\paragraph{Charge state control}
We start by presenting optical emission spectra recorded from typical individual Q-Dots and Q-Posts under weak optical pumping ($\sim 100$~nW) as a function of the applied SAW power, $P_{\rm SAW}$. This data is encoded in a greyscale representation in Fig.~\ref{Hysterese}. Furthermore, we compare the spectra for different sweep-directions of the SAW power. The left (right) panels  show the detected emission for increasing (decreasing) $P_{\rm SAW}$ which we refer to as the up-sweep (down-sweep).\\
For the up-sweep of the Q-Post shown in (a) we observe for very low SAW-powers two dominant emission lines arising from singly negatively charged excitons $(1X^-=2e+1h)$. Details on the attribution of all observed lines can be found in Ref. \cite{Voelk_NanoLett}. At $P_{\rm SAW}\simeq \pm0{\rm ~dBm}$, the power level at which the Matrix-QW emission starts to quench and acoustic charge conveyance set in, we observe a distinct switching behavior to emission of the charge neutral single exciton$(1X^0=1e+1h)$ shifted by $\sim 7$ meV towards higher energies and remaining the dominant emission for further increasing $P_{\rm SAW}$. When $P_{\rm SAW}$ is subsequentially decreased during the down-sweep shown in (b), we observe an almost identical switching behavior with exactly the same threshold level at $P_{\rm SAW} \simeq +0.50$~dBm. \\  
The observed switching behavior is significantly different for a conventional self-assembled Q-Dot as shown in panels (c) and (d): The first striking observation is the splitting into three characteristic emission lines at high SAW power levels. These lines are attributed to the neutral exciton ($1X^0$) and biexciton ($2X^0=2e+2h$) and the positively charged exciton ($1X^+=1e+2h$). Since these occupancy states cannot exist simultaneously, their observation in a time-integrated spectrum unambiguously indicates that the net charge state of the Q-Dot is not stable after the onset of acoustic charge conveyance in contrast to the Q-Post for which the neutral configurations $1X^0$ and $2X^0$ are favored. Furthermore, the switching behavior of the Q-Dot is highly hysteretic: the threshold during the SAW up-sweep occurs at $P_{\rm SAW} \simeq +18$~dBm which is approximately one order of magnitude larger than during the down-sweep which is found at $P_{\rm SAW} \simeq +9$~dBm. At the highest SAW power levels, we also observe a characteristic spectral broadening of the emission lines due to a dynamic modulation of the exciton energies by the SAW fields~\cite{Gell_APL,Metcalfe_PRL}.\\   
The dissimilar switching behavior of Q-Dot and Q-Post can be readily understood by taking into account the two-dimensional systems to which these nanostructures are coupled to and within which acoustic charge conveyance occurs~\cite{Voelk_NanoLett}: For the Q-Dot, charge separation and conveyance occur within the underlying thin WL while for the Q-Post, these processes occur within the 23~nm wide Matrix-QW. For the Matrix-QW emission, we observe the well-known dynamic SAW modulation as described in detail in Refs.~\cite{Alsina,Voelk_APL}. The resulting efficient separation of electrons and holes gives rise to an inherently sequential injection of the two carrier species: At a given time, electrons (holes) are present at the Q-Post and half a SAW period later the opposite charge, holes (electrons). For the Q-Post, this sequential injection largely suppresses the formation of charged exciton complexes and enhances the generation probability of neutral species. To understand the significantly different behavior observed for Q-Dots, we have to take into account the carrier mobility in a QW. It is highly dependent on the width of the QW within which transport occurs~\cite{Sakaki} and thus, dissociation of photogenerated charges and acoustic charge conveyance are hampered within the WL. In particular, the mobility of holes is highly reduced which even localize and get trapped by randomly formed potential fluctuations~\cite{Fuhrmann_JAP,Babinski} leading to the formation of 'natural' Q-Dots while transport of electrons out of the generation area is less affected. This misbalance of the mobilities gives rise to a large net increase of the hole density at the position of the Q-Dot which results in an efficient generation of the observed $1X^+$ state~\cite{Voelk_NanoLett}. During the down-sweep in the bistable region of the hysteresis, the net increase of the $2X^0$ intensity over that of the $1X^0$ points towards efficient hole capture on timescales even faster than the radiative cascade from $2X^0\rightarrow1X^0\rightarrow0$ due to a complete break down of hole charge conveyance.
\paragraph{Enhanced carrier injection}
Another characteristic feature is the net increase of the overall emission intensity at the onset of acoustic charge conveyance for both systems. 
\begin{figure}[htb]%

  \includegraphics*[width=0.9\columnwidth]{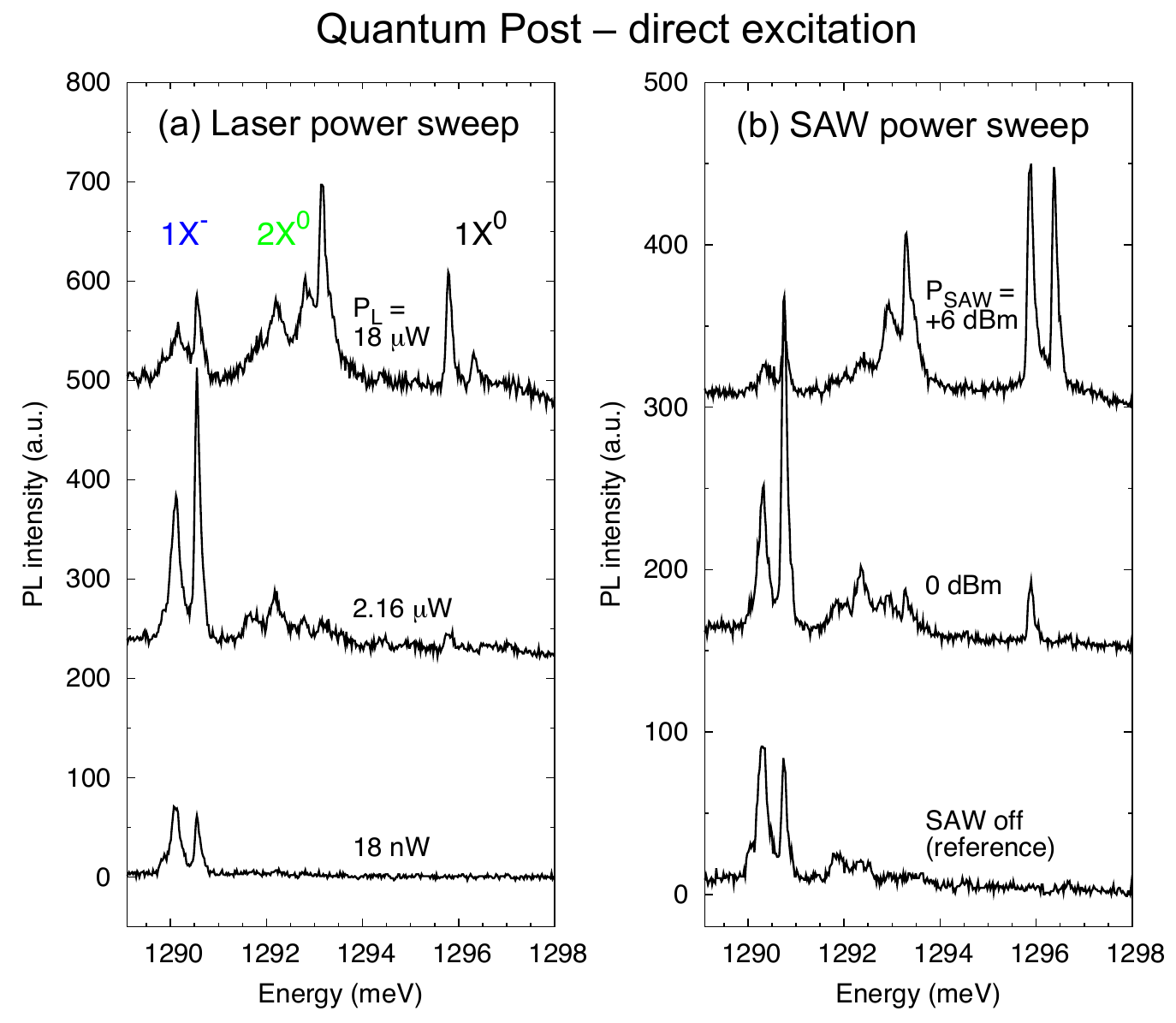}%

  \caption[]{%
  (color online) (a) Laser power dependent emission of a single Q-Post showing the emergence of biexcitonic lines at high pump power. (b) SAW power dependent spectra recorded at low optical pump intensities showing an overall increase of the integrated intensity and switching from charged to neutral exciton lines at high SAW power.}
    \label{Power}
\end{figure}
In panels (a) and (b) of Fig.~\ref{Power}, we compare spectra recorded from a single Q-Post under direct optical excitation as a function of the applied optical pump power (without SAW) and as a function of $P_{\rm SAW}$ (at constant optical pump power), respectively. For low optical pump power, we observe the aforementioned dominant emission of $1X^-$. For increasing pump power, the generation of neutral and charged biexcitons is favored giving rise to additional emission of $1X^0$ resulting from the intermediate level of the biexciton-exciton cascade. In a second experiment, we kept the optical pump power constant at a low level at which only single exciton species are formed [Fig.~\ref{Power} (b)]. Without a SAW applied, we detect strong emission of $1X^-$. For $P_{\rm SAW}=\pm0$~dBm close to the onset of charge conveyance, we observe an overall increase of the intensity of the $1X^-$ lines and additional emission of the initially suppressed $2X^0$ and $1X^0$ states and the integrated intensity of the Q-Post emission approximately doubles. In this power range, diffusion driven carrier separation dominates which is less efficient for holes compared to electrons \cite{Garcia}. Therefore, additional electrons from within the generation area (laser spot diameter) are accumulated at the position of the Q-Post giving rise to the observed increase of the emission intensity. This effect is even more pronounced for Q-Dots [cf. Fig.~\ref{Hysterese}] due to the significantly lower hole mobility in the thin WL~\cite{Schuelein_new}. For further increasing $P_{\rm SAW}$ to $+6$~dBm, the switching of the emission pattern to the $1X^0$ and $2X^0$ lines occurs and the integrated intensity decreases due to more efficient transport of holes out of the generation area i.e. the position of the Q-Post.

\subsection{Remote injection}
\begin{figure}[htb]%
% \sidecaption
  \includegraphics*[width=0.9\columnwidth]{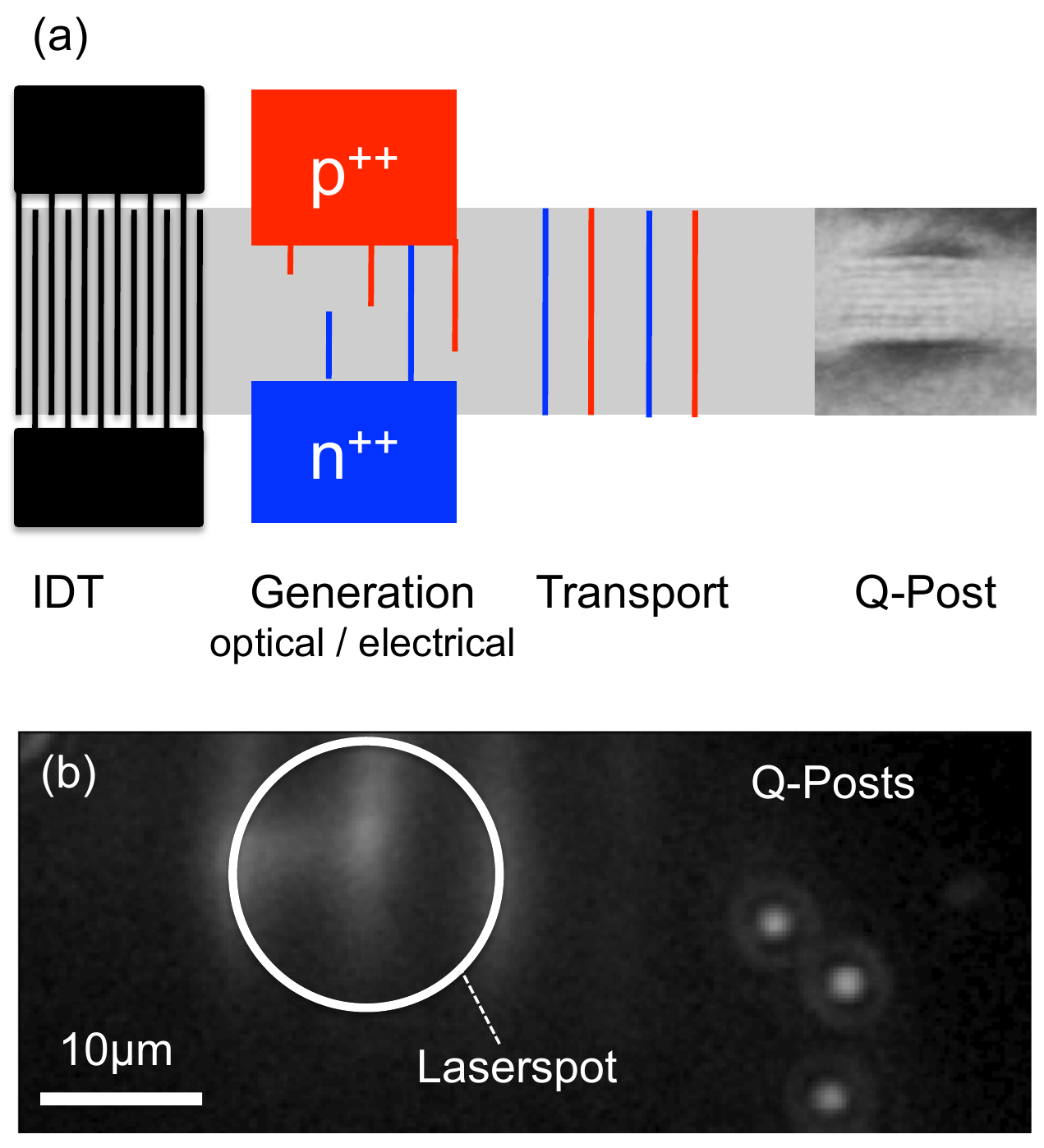}%
  \caption[]{%
  (color online) (a) All-electrical, SAW-triggered single photon source. The electrons (blue) and holes (red) are injected by a lateral \emph{p-i-n} junction and transported by a SAW as indicated by the vertical lines. (b) PL-image showing emission of the Matrix-QW at the laserspot position and of remotely located, individual Q-Posts into which carriers are acoustically injected.}
   \label{Remote}
   \end{figure}

All experimental data presented to this point was recorded under direct optical excitation [c.f. Fig.~\ref{directandremote} (a)]. In order to implement the scheme proposed by Wiele \textit{et al.}~\cite{Wiele}, we have to \emph{spatially separate} the carrier generation area and the quantum emitter as shown in Figs.~\ref{directandremote} (b) and \ref{Remote} (a) for optical and electrical carrier generation, respectively.
In order to confirm acoustically mediated carrier injection into a remote Q-Post, we illuminate our sample with a slightly widened laser spot (diameter $\sim 15~\mu\mathrm{m}$) and image the emission (Matrix-QW and Q-Posts) directly on our CCD detector. A typical PL-image is shown in Fig.~\ref{Remote} (b) for  $P_{\rm SAW}=+12$~dBm with the IDT positioned on the left outside the field of view. At the position of the laserspot, we detect a stripe pattern arising from the dynamic suppression of the Matrix-QW emission~\cite{Rocke,Alsina}. Along the SAW propagation direction on the righthand side, outside of the optical generation area, we resolve clearly point-like emission arising from \emph{individual} Q-Posts into which the photogenerated electron-hole pairs are acoustically injected as proposed~\cite{Wiele} and recently demonstrated for interface fluctuation Q-Dots~\cite{Couto}. This is further confirmed by the observation of narrow line emission comparable to that in  Figs. \ref{Hysterese} and \ref{Power}.
\section{Conclusions and future directions}

In summary, we have shown that a SAW provides a versatile high frequency tuning mechanism for Q-Dot and Q-Post occupancy states and emission energies. Furthermore, we demonstrated that an acoustically triggered single photon source can be readily implemented using these high-quality quantum emitters. To fully exploit the potential, all electrical operation could be realized by using lateral \emph{p-i-n} junctions~\cite{Foden,Gell_APL-pn,Stavarche} as shown schematically in Fig.~\ref{Remote} (a). The high frequency deterministic generation of the neutral biexciton state could be particularly attractive for the realization of an entangled photon pair source for which such high preparation fidelities are crucially required and difficult to achieve using alternative schemes such as Coulomb-blockade~\cite{Benson}. Future research could be directed towards the spin degree of freedom \cite{Sogawa,Stotz,book} in single or coupled Q-Dot systems in planar~\cite{Krenner-05,Krenner-NJP,Kiravittaya} and nanowire architectures~\cite{Kinzel_NanoLett,Weert,Kouwen} or in nanophotonic cavities~\cite{Fuhrmann:11}.

\section{Acknowledgements}
This work was supported by DFG via the "Nanosystems Initiative Munich" (NIM), the Emmy Noether Program, SFB491 and SPP1285, BMBF via nanoQUIT, NSF via NIRT grant CCF-0507295, DARPA SEEDLING and the Alexander-von-Humboldt Foundation.

\end{document}